\documentclass[aps, twocolumn, english, superscriptaddress, nofootinbib, preprintnumbers]{revtex4-1}

\usepackage{color}
\usepackage{babel}
\usepackage{amsmath}
\usepackage{graphicx}
\usepackage{amssymb}
\usepackage{soul}
\usepackage{float}
\usepackage{subfig}
\usepackage{hyperref}



\newcommand{\Hil}{\mathcal{H}}
\newcommand{\id}{\mathbb{I}}

\begin{document}

\title{Emergent de Sitter epoch of the quantum Cosmos}

\author{Mehdi Assanioussi}
\email{mehdi.assanioussi@fuw.edu.pl}
	\affiliation{Faculty of Physics, University of Warsaw, Pasteura 5, PL-02-093 Warszawa, Poland.}

\author{Andrea Dapor}
\email{andrea.dapor@gravity.fau.de}
\affiliation{Institute for Quantum Gravity, Friedrich-Alexander University Erlangen-N\"urnberg, Staudstra\ss e 7, 91058 Erlangen, Germany.}

\author{Klaus Liegener}
\email{klaus.liegener@gravity.fau.de}
\affiliation{Institute for Quantum Gravity, Friedrich-Alexander University Erlangen-N\"urnberg, Staudstra\ss e 7, 91058 Erlangen, Germany.}

\author{Tomasz Paw{\l}owski}
	\email{tomasz.pawlowski@fuw.edu.pl} 
	\affiliation{Center for Theoretical Physics, Polish Academy of Sciences, Al. Lotnik\'ow 32/46, 02-668 Warsaw, Poland.}
	\affiliation{Faculty of Physics, University of Warsaw, Pasteura 5, PL-02-093 Warszawa, Poland.}

\date{\today{}}

\begin{abstract}
\noindent The quantum nature of the Big Bang is reexamined in the framework of Loop Quantum Cosmology. The strict application of a regularization procedure to the Hamiltonian, originally developed for the Hamiltonian in loop quantum gravity, leads to a qualitative modification of the bounce paradigm. Quantum gravity effects still lead to a quantum bounce connecting deterministically large classical Universes. However, the evolution features a large epoch of de Sitter Universe, with emergent cosmological constant of Planckian order, smoothly transiting into a spatially flat expanding Universe.
\end{abstract}

\maketitle

Observations \cite{P2015} show that our current knowledge of physical processes in the early Universe is far from complete, indicating the necessity for a deeper understanding of the interactions between matter and spacetime in strong field regime. It is expected, that an accurate description of the early Universe would require taking into account both the relativistic nature of gravity and the quantum nature of physical fields. A promising approach to construct such description is Loop Quantum Gravity (LQG) \cite{t-book}. LQG quantization provides a quantum representation which is different from the Fock representation of standard quantum field theory, but compatible with the background independence principle dictated by general relativity. Advances in LQG, and especially in Loop Quantum Cosmology (LQC) \cite{b-liv, *A11} which applies LQG quantization methods to cosmological spacetimes, have opened a promising avenue to close the gap in our understanding of the interactions between geometry and quantum matter, by probing the quantum geometry effects in the Planck regime, and they have the potential to leave observational signatures \cite{agn-pert}. In particular, the LQC model of flat Friedman-Lema\^{i}tre-Robertson-Walker (FLRW) Universe led to the replacement of the big bang initial singularity by a bounce, connecting two (semi-)classical FLRW spacetimes in a deterministic manner \cite{aps-prl}. In this letter, we discuss a new evolution scenario within LQC where the pre-bounce geometry could be described on the effective level by a de Sitter spacetime, with a Planck scale positive cosmological constant.

The LQC framework, like the full LQG program, is based on Dirac's canonical quantization of the system in its Hamiltonian formulation. The canonical variables are chosen to be holonomies (parallel transports) of the so-called Ashtekar connections along curves and fluxes of densitized triads across surfaces. The LQG quantization provides a Hilbert space which identifies with the space of square integrable functions on the configuration space of (generalized) three-dimensional spatial Ashtekar connections. The observables, such as the Hamiltonian, are then implemented as quantum operators acting in this Hilbert space. This implementation requires a regularization of the classical observable in order to express it in terms of holonomies and fluxes. In the literature several regularization proposals exist \cite{QSD1, NSC, CO}. In the current letter, we implement the first one devised for full LQG \cite{QSD1}, which is the main point of departure from mainstream LQC.
Specifically, in terms of Ashtekar variables the classical gravitational Hamiltonian takes the form 
\begin{equation}\label{eq:hamgr}\begin{split}
	H_{gr}(N) =& N(H_E + H_L) \\
		:=& \frac{N}{16\pi G} \int d^3x \frac{ \epsilon^{ij}{}_k E^a_i E^b_j F^k_{ab}}{\sqrt{{\rm det}(E)}}  \\
		&- N \frac{1+\gamma^2}{8\pi G} \int d^3x \frac{ E^a_iE^b_jK^i_{[a}K^j_{b]}}{\sqrt{{\rm det}(E)}} .
\end{split}\end{equation}
where $N$ is the lapse function, $\gamma$ is the Immirzi-Barbero parameter, $F^k_{ab}$ is a curvature of the Ashtekar connection $A^i_a$, $E^a_i$ is the densitized triad conjugate to $A^i_a$ and $K^i_a$ is the extrinsic curvature. 

In classical flat cosmology we have
\begin{equation}\label{eq:clcos}
	\gamma K^i_a = A^i_a\;\;\;{\rm and}\;\;\; 2\gamma^2 K^i_{[a}K^j_{b]}= \epsilon^{ij}{}_k F^k_{ab} .
\end{equation}
Hence $H_{gr}$ becomes proportional to $H_E$ and its quantisation describes the standard dynamics in LQC \cite{aps-imp,pa-posL}.
The regularization we implement here, is based on the identity \cite{QSD1} 
\begin{equation}\label{eq:K}
	K^i_a = \frac{1}{8\pi G\gamma^3}\{A^i_a, \{H_E,V\}\} ,
\end{equation}
valid in full general relativity. Under the choices \eqref{eq:clcos} and \eqref{eq:K}, the Hamiltonian in \eqref{eq:hamgr} becomes respectively
\begin{equation}\label{oldH}
	H_{gr}(N) = -\frac{3 N}{8\pi G\gamma^2\Delta} V \sin^2(b) ,
\end{equation}
and
\begin{equation}\label{ourH}
	H_{gr}(N) = \frac{3 N}{8\pi G\Delta} V \left(\sin^2(b) - \frac{1+\gamma^2}{4\gamma^2} \sin^2(2b)\right) ,
\end{equation}
where $V$ and $b$ are canonical variables in cosmology (see below) and $\Delta=2\pi \sqrt{3}\gamma G\hbar\approx 2.61 l_{Pl}^2$, with $l_{Pl}$ the Planck length and $\gamma=0.24$, is the so called LQC area gap \cite{abl-lqc}).
The Hamiltonian \eqref{ourH} has been considered already in \cite{YDM09}, however the resulting quantum dynamics has not been analyzed. As we show in this letter, the pre-bounce evolution of the Universe induced by \eqref{ourH} differs significantly from that of standard LQC [which uses \eqref{oldH}].

It is worth noting that differences akin to those studied here can occur even in simple quantum mechanical systems, where the observations allow to single out the ``correct'' (physically preferred) dynamics. Similarly, in cosmology the different possibilities have to be pursued in order to identify possible observational imprints of each choice and to identify the ``right'' one. 
We do so using as an example the model of a flat FLRW Universe filled with massless scalar field. Note that the quantization of geometry and the qualitative results would hold with other standard matter content. In this case, the matter degrees of freedom are the field value and its conjugate momentum, while the geometry degrees of freedom are captured by a canonical pair of global variables: oriented volume $V$ of the chosen compact region ${\cal V}$ (playing the role of the infrared regulator) and its conjugate momentum proportional to Hubble rate $H_r$. For convenience, we choose the same pair of variables $(v,b)$ as in LQC \cite{aps-imp}, such that $V = 2\pi\gamma\sqrt{\Delta}G\hbar |v| =: \alpha |v|$ and $\{v,b\} = 2/\hbar$.

The quantization of this theory is obtained following LQG methods \cite{abl-lqc}: for the geometry, one promotes $v$ and $e^{i \lambda b}$ 
(for $\lambda \in \mathbb R$)\footnote{These functions are sufficient to encode relevant subalgebra of holonomies and fluxes.} -- not $b$ -- 
to operators on the Hilbert space of square-integrable functions on the Bohr compactification of the real line, $\mathcal H_{\rm gr} := L_2(\mathbb R_{\text{Bohr}}, d\mu)$. As for matter, one uses standard Schr\"odinger representation $\mathcal H_{\phi} = L_2(\mathbb R, d\phi)$, on which $\hat \phi = \phi$ and $\hat p_\phi = i\hbar \partial_\phi$. The Hamiltonian constraint
\begin{equation}\label{eq:ham}
	H(N) = H_\phi(N) + H_{gr}(N) :=N\frac{p_\phi^2}{2V}+H_{gr}(N)=0 ,
\end{equation}
with $H_{gr}(N)$ of \eqref{ourH}, is promoted to a constraint operator acting on $\Hil_{\rm kin}= \Hil_{\rm gr} \otimes \Hil_{\phi}$.
After a convenient choice of lapse $N=2V$, the quantum constraint equation corresponding to \eqref{eq:ham} reads
\begin{equation}\label{eq:ev}
\begin{split}
  &\hat{p}_{\phi}^2\Psi(v,\phi) = \hbar^2\Theta\otimes\id\;\Psi(v,\phi), \quad \Theta:\Hil_{\rm gr} \to \Hil_{\rm gr} , \\
	&\Theta = \frac{3\pi G}{4} \gamma^2 \left[-s f_8(v) {\cal N}^8 - s f_{-8}(v) {\cal N}^{-8}\right. \\
	&+ \left.f_4(v) {\cal N}^4 + 2 (s-1) f_0(v) \id + f_{-4}(v) {\cal N}^{-4}\right]
\end{split}\end{equation}
where $f_a(v) = \sqrt{|v(v+a)|}|v+a/2|$, $[{\cal N}^k\psi](v)=\psi(v+k)$ and $s=(1 + \gamma^2)/(4\gamma^2)$. Unlike the standard LQC where the $\Theta$ is a difference operator of the $2$nd order, in this case $\Theta$ is now a difference operator of the $4$th order.

The last step of Dirac program, that is the construction of the physical Hilbert space, is performed systematically via the so-called group averaging procedure \cite{almmt-ga}. For the considered model, this is mathematically equivalent to a deparametrized system where the dynamical field $\phi$ is interpreted as an internal clock parametrizing the evolution for a free ``geometry field''. The operator $\Theta$ becomes then the square of a physical Hamiltonian. Taking the positive frequency superselection sector (by analogy with Klein-Gordon equation) one arrives to a Schr\"odinger evolution equation
\begin{equation}
	-i\partial_{\phi}\Psi(v,\phi) = \sqrt{|\Theta|}\Psi(v,\phi) ,
\end{equation}
where $|\Theta|$ is the positive part of (a self-adjoint extension of) the operator $\Theta$. The physical Hilbert space is then the proper subspace of $\Hil_{\rm gr}$ defined by the spectral decomposition of $|\Theta|$.

Since $\Theta$ is a $4$th order difference operator which preserves the lattices $\mathcal{L}_{\epsilon} = \epsilon+4\mathbb{Z}$, and the parity reflection $v\mapsto -v$ is a gauge transformation, it follows that the physical Hilbert space is divided into superselection sectors of (anti)symmetric functions supported on particular lattices. For convenience, we focus on the sector of symmetric functions supported on $v\in 4\mathbb{Z}$, which makes $b$ a periodic coorindate.\footnote{%
  The choice of other lattices, despite giving slightly more complicated mathematical structure of the solutions, leads to very similar physical results.} 
On a single sector, $\Theta$ generates the dynamics of a $2$nd order system. To see this, it is convenient to perform a transformation to the $b$-representation, which brings the operator to the form
\begin{equation}\label{theta}
	\Theta = 12\pi G \gamma^2[ (\sin(b)\partial_b)^2 - s(\sin(2b)\partial_b)^2] .
\end{equation}
This expression is similar to the one for the evolution operator $\Theta_\Lambda$ in the FLRW model with a positive cosmological constant \cite{pa-posL},
\begin{equation}
	\Theta_\Lambda = -12\pi G \gamma^2[ (\sin(b)\partial_b)^2 - (b_\Lambda\partial_b)^2] ,
\end{equation}
with $b_\Lambda \in \mathbb{R}$. The difference with \eqref{theta} is that $b_\Lambda$ is replaced by a function on the phase space. However, this similarity allows us to employ the methodology developed in \cite{pa-posL} to analyze the operator $\Theta$.
The analysis shows that the operator $\Theta$ admits a $U(1)$ family of self-adjoint extensions $\{\Theta_\beta\}_{\beta\in [0,2\pi)}$, each generating a unique unitary evolution. However, the numerical investigations show that for sharply peaked states the results are largely insensitive to the choice of self-adjoint extension.

In the $v$-representation, an eigen-function $e_{k,\beta}$ corresponding to an eigenvalue $\omega^2 = 12\pi G k^2$ has no simple analytic form; however it has the following asymptotic behaviour
\begin{equation}\label{eq:e-as}\begin{split} 
	e_{k,\beta}(v) &= \frac{4}{\sqrt{2\pi v}} \cos(ik\ln|v|+\varphi_{\beta}(k)) \\
	&+ \frac{R_{\beta}(k)}{v} \cos(i\Omega|v|+\chi_{\beta}(k)) 
	+ O(|v|^{-3/2}) ,
\end{split}\end{equation}
where $\cos(4\Omega) = (2s-1)/2s$, $R_{\beta}(k)$ is a positive amplitude and $\varphi_{\beta}, \chi_{\beta}$ are phase shifts.
This expression allows us to deduce the properties of the physical states for large volume. In particular, the first term has the same form as in standard LQC and it indicates the presence of two epochs (expanding and contracting) of a FLRW classical evolution. Similarly, (by comparing the asymptotics), the second term indicates two epochs (again contracting and expanding) of a scalar-de Sitter Universe with a cosmological constant 
\cite{pa-posL}
\begin{equation}\label{eq:Lambda}
	\Lambda = 3/(\Delta (1+\gamma^2)^2) \approx 1.03\ l_{Pl}^{-2} .
\end{equation}
(where we take $\gamma=0.24$), again indicating the precence of expanding and contracting epoch. Note however, that the cosmological constant is not put in by hand. Instead, it emerges from quantum geometry effects. 

Having the eigenstates of $\Theta$, we can study  the evolution of physical states. For each extension $\Theta_\beta$, the physical states have the form
\begin{equation}\label{eq:state}
	\Psi(v,\phi) = \int_0^{+\infty} dk \tilde{\Psi}(k) e_{k,\beta}(v) e^{i\omega(k)\phi} ,
\end{equation}
where $\tilde{\Psi}$ is the spectral profile and $\omega(k) = \sqrt{12\pi G} k$. To extract meaningful physical information regarding the dynamics of the Universe, one has to select a suitable set of observables. We use in particular:
	
$\bullet$ the compactified volume: ${\theta}_K := \arctan(|v|/K)$ for a conveniently chosen positive parameter $K$. Unlike the actual volume operator, which would map certain states outside of the Hilbert space, this operator is bounded with $\langle\theta_K\rangle = \pi/2$ corresponding to infinite volume.
	
$\bullet$ the matter energy density: ${\hat\rho} := V^{-1}\Theta V^{-1}$ which is a bounded operator with a continuous part of the spectrum ${\rm Sp}_{\rm cont}(\hat{\rho}) = [0,\rho_c]$, where
\begin{align}\label{rhoc}
	\rho_c=3/(32\pi G \Delta \gamma^2(1+\gamma^2))\approx 0.19 \rho_{Pl} ,
\end{align}
in contrast with the value $0.41 \rho_{Pl}$ obtained in standard LQC ($\gamma=0.24$).

$\bullet$ the Hubble rate: ${H}_r = \frac{i}{6} [V,V^{-1/2}\Theta V^{-1/2}]$, again a bounded operator.

These observables can be used to numerically probe the dynamics of Gaussian states 
$\tilde{\Psi}(k) \propto \exp(-(k-k^\star)^2/4\sigma^2)$ peaked on $p_{\phi} = \sqrt{12\pi G} k^\star$ with variance $\Delta p_{\phi} = \sqrt{12\pi G}\sigma$. In the actual numerical investigations, we consider a state sharply peaked on a classical configuration and evolve it backward in physical time. We then calculate the backward evolution of the expectation values and variances of the observables listed above.
It turns out that the quantum trajectories of these quantities are to high precision independent of the choice of the self-adjoint extension.
\begin{figure*}
	\begin{centering}
	\subfloat[]{\includegraphics[width=0.45\textwidth,height=0.34\textwidth]{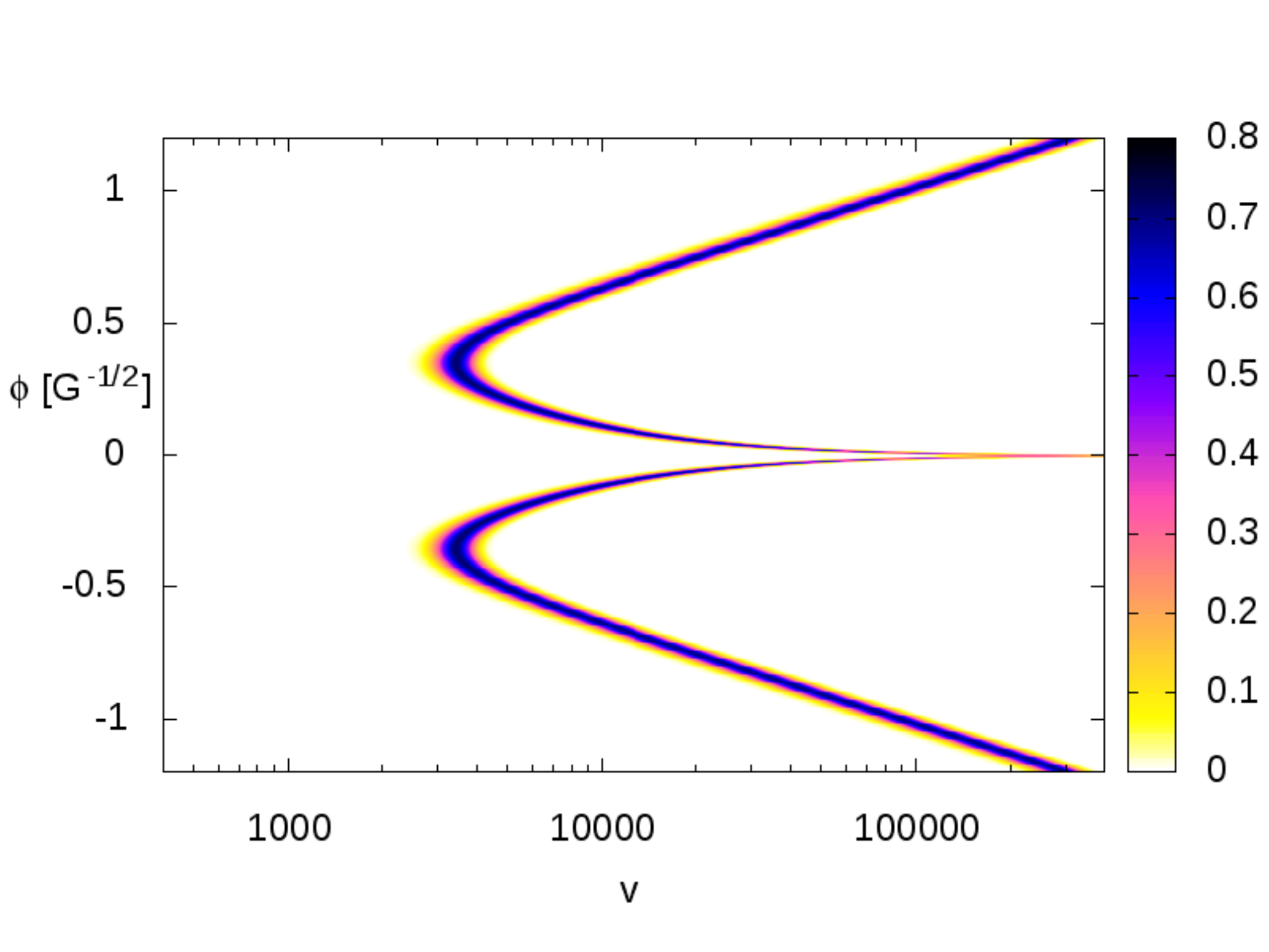}\label{figs1a}}
	\hspace{0,5cm}
	\subfloat[]{\includegraphics[width=0.45\textwidth]{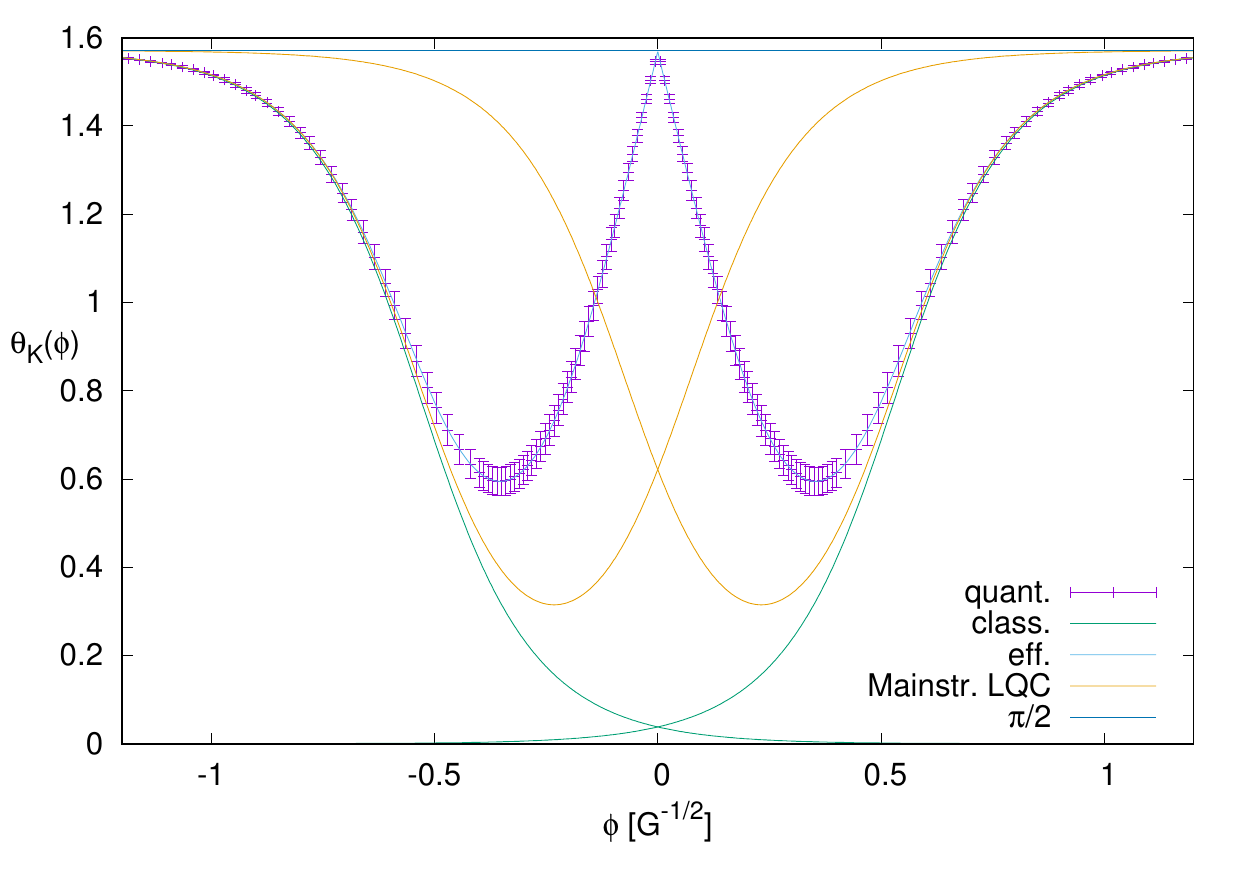}\label{figs1b}} \\
	\subfloat[]{\includegraphics[width=0.45\textwidth]{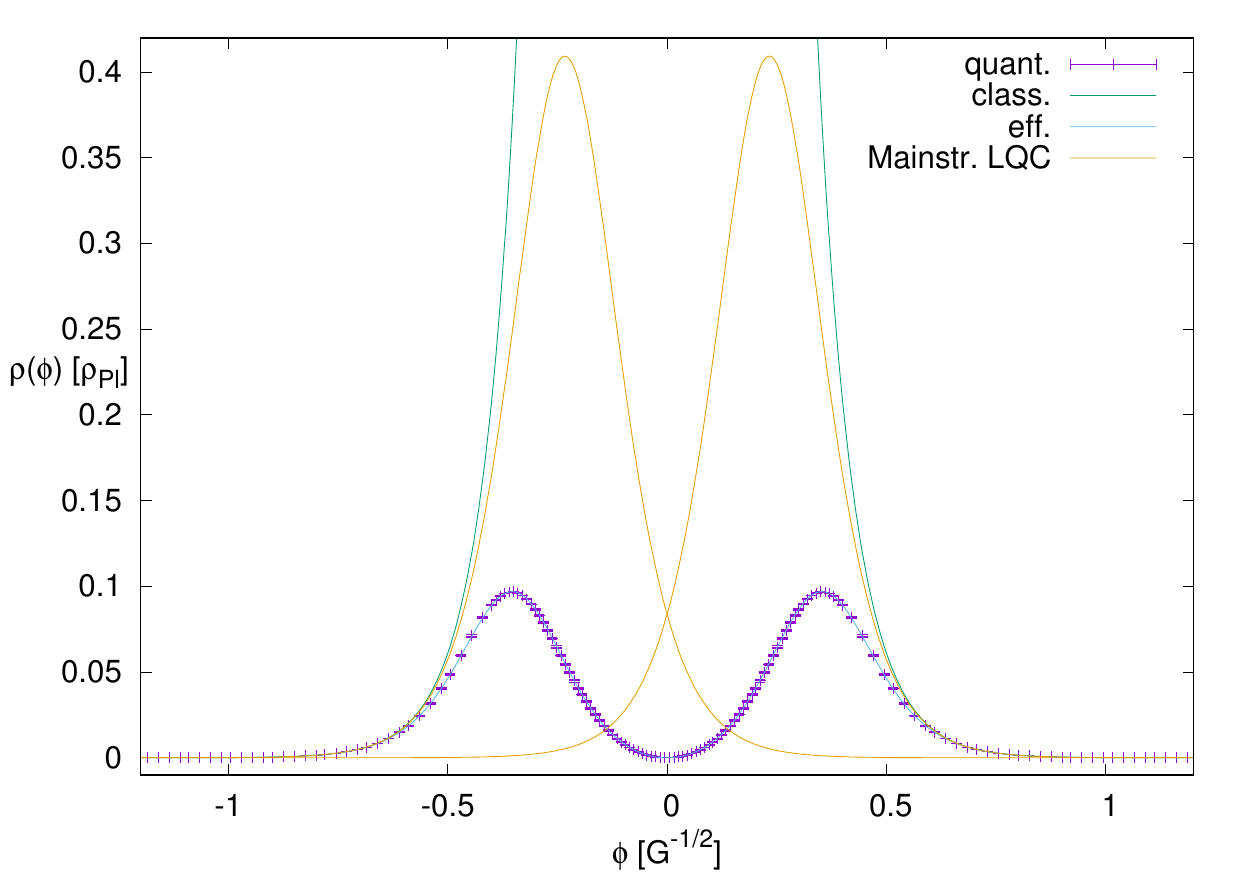}\label{figs1c}}
	\hspace{0,5cm}
	\subfloat[]{\includegraphics[width=0.45\textwidth]{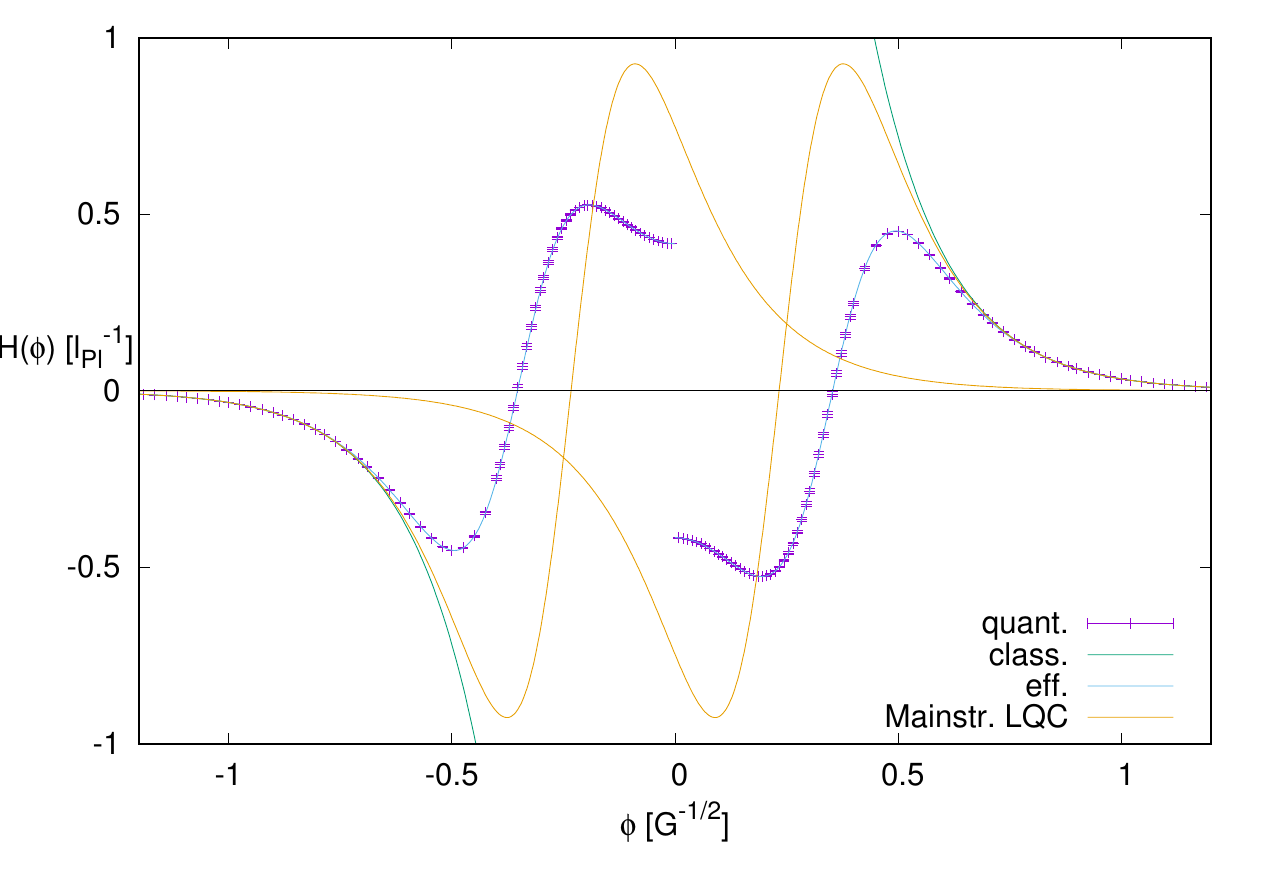}\label{figs1d}}
	\end{centering}
\caption{\small \raggedright The map of the physical state $(a)$ {on a $v-\phi$} plane [where the volume $V\approx 2.41 |v| \ell_{\rm Pl}^3\ $], and quantum trajectories of the observables: {compactified volume} $\theta_{K=5\cdot 10^3}$ $(b)$, {matter energy density} $\rho_{\phi}$ $(c)$ and {Hubble rate} $H_r$ $(d)$ of the Gaussian state peaked on $p_{\phi} = 5.05\cdot 10^3 G^{1/2}$ with relative spread in $\Delta p_{\phi}$ of about $0.05$. {The genuine quantum trajectories of the investigated model (purple error bars) are compared against the predictions of the effective dynamics generated by Hamiltonian \eqref{Effec-Ham} (blue lines) and against the classical GR (green lines) and mainstream LQC  effective trajectories (yellow lines), to which the quantum one converges in the asymptotic past/future. While both mainstream LQC trajectories feature a single bounce (each) at (respectively) $\phi\approx\pm 0.25 G^{-1/2}$, for the trajectories obtained with the Hamiltonian we investigate \eqref{theta} we observe two bounces at $\phi\approx \pm 0.35 G^{-1/2}$ separated by a a transition point from future to past conformal infinity at $\phi = 0$, where the matter energy density reaches zero and the volume $V$ reaches infinity.} The Planck units $\rho_{\rm Pl}$ and $\ell_{\rm Pl}$ are defined respectively as $(G^{2}\hbar)^{-1}$ and $(G\hbar)^{1/2}$. 
The departure from mainstream LQC lasts only {about $1.2 G^{-1/2}$} in relational time $\phi$, but from each bounce it takes infinite cosmic time to reach the transition at $\phi = 0$.}
\label{figs1}
\end{figure*}
An example ($\beta=0$) of the results is presented in fig.\ref{figs1}. The state sharply peaked at some initial $\phi$ remains so throughout the evolution, but its evolution is much richer and involves distinct epochs. {In backward evolution, when starting with the semiclassical initial data (state at given time sharply peaked in relevant observables) corresponding to a large (low energy density) expanding Universe, we observe (see fig.~\ref{figs1}) that the quantum trajectory follows the predictions of GR till matter energy density reaches Planckian 
order, where (similarly to the standard LQC) gravitational forces become repulsive and cause the bounce. The energy density is lower than in mainstream LQC but remains of the same order \eqref{rhoc}. Past the bounce, instead of entering the 
classical trajectory again the Universe enters a contracting de Sitter epoch, where the quantum geometry effects qualitatively manifest themselves as the effective cosmological constant of Planckian order \eqref{eq:Lambda}. This epoch lasts till the matter energy density reaches zero and the volume $V$ reaches infinity, at which point there occurrs a transition through past conformal infinity (past scri) into a qualitatively symmetric expanding de Sitter epoch and another bounce at the critical energy density \eqref{rhoc}. Past the second bounce the quantum trajectory quickly starts to follow predictions of GR without cosmological constant.}

This new scenario can be further analyzed if one resorts to the effective classical theory, in which the effective Hamiltonian is obtained by replacing operators with classical functions. Under this rule, one finds
\begin{equation}\label{Effec-Ham}
H_{\text{eff}} = \frac{p_\phi^2}{2\alpha v} - \frac{3\pi G\hbar^2}{2\alpha} v \sin^2(b) \left[1 - (1 + \gamma^2) \sin^2(b)\right] ,
\end{equation}
where now the lapse is set to $N=1$ and as before $\alpha=2\pi \gamma \sqrt{\Delta} G \hbar$. 
Solving $H_{\text{eff}} = 0$ for the matter energy-density yields 
\begin{align}
\rho := \frac{p_\phi^2}{2\alpha^2 v^2} = \frac{3\pi G}{2\alpha^2} \sin^2(b) [1 - (1 + \gamma^2) \sin^2(b)] . 
\end{align}
This shows that $\rho$ is bounded, and its maximum value is $\rho_c$, that is the critical matter energy density given by \eqref{rhoc}. We can use $H_{\text{eff}}$ to compute Hamilton's equations for $\dot v$ and $\dot b$ for initial conditions set at the bounce, where we choose $b\in [-\pi/2,\pi/2]$. The bounce is identified by $\dot v = 0$, which can be solved for the matter energy density to find $\rho = \rho_c$. As for the value of $b$ and $v$ at the bounce, one finds $b_B = \pm \arcsin(1/\sqrt{2(1 + \gamma^2)}),$ $\ v_B = |p_\phi| \sqrt{4(1+\gamma^2)/3\pi G}$.
The only free parameters are $p_\phi$ (which is a constant of motion) and the sign of $b_B$. Hamilton's equations can now be numerically integrated, and as shown in fig.\ref{figs1}, we find an agreement better than $0.1\sigma$ between the effective dynamics and the quantum dynamics discussed previously.
Contrary to standard LQC, the analysis of the large-volume behavior leads to two possible scenarios:
i) the Universe starts in the far past with $b = 0$ and evolves into the far future with $b = |b_o|$; ii) the Universe starts in the far past with $b = -|b_o|$ and evolves into the far future with $b = 0$, where $b_o$ is given by the condition $\sin^2(b_o)(1+\gamma^2) = 1$.
What controls which of the two cases is realized, is the sign of $b_B$: if $b_B > 0$, then we are in the first case; if $b_B < 0$, then we are in the second case. Expanding $H_{\text{eff}}$ around $b=0$ and $b=\pm b_o$, we find respectively
\begin{subequations}
\begin{align}
	H_{\text{eff}} &\stackrel{b \rightarrow 0}{\longrightarrow} \dfrac{p_\phi^2}{2\alpha v} - \dfrac{3\pi G\hbar^2}{2\alpha} v b^2,\\ 
	H_{\text{eff}} &\stackrel{b \rightarrow \pm b_o}{\longrightarrow} \dfrac{p_\phi^2}{2\alpha v} - \sqrt{\dfrac{3\Lambda}{4}} \hbar v (b_o \mp b)
\end{align}
\end{subequations}
where $\Lambda$ is defined in \eqref{eq:Lambda}. Friedmann equation follows from Hamilton's equations in these two regimes. The results are
\begin{align}
\left(\frac{\dot v}{3v}\right)^2 = \left\{
\begin{array}{ll}
\dfrac{8\pi G}{3} \rho & \text{if} \ b \rightarrow 0
\\
\\
\dfrac{\Lambda}{3} & \text{if} \ b \rightarrow \pm b_o
\end{array}
\right.
\end{align}

\begin{figure*}[t]
\centering
	\subfloat[]{\includegraphics[width=7cm, height=5cm]{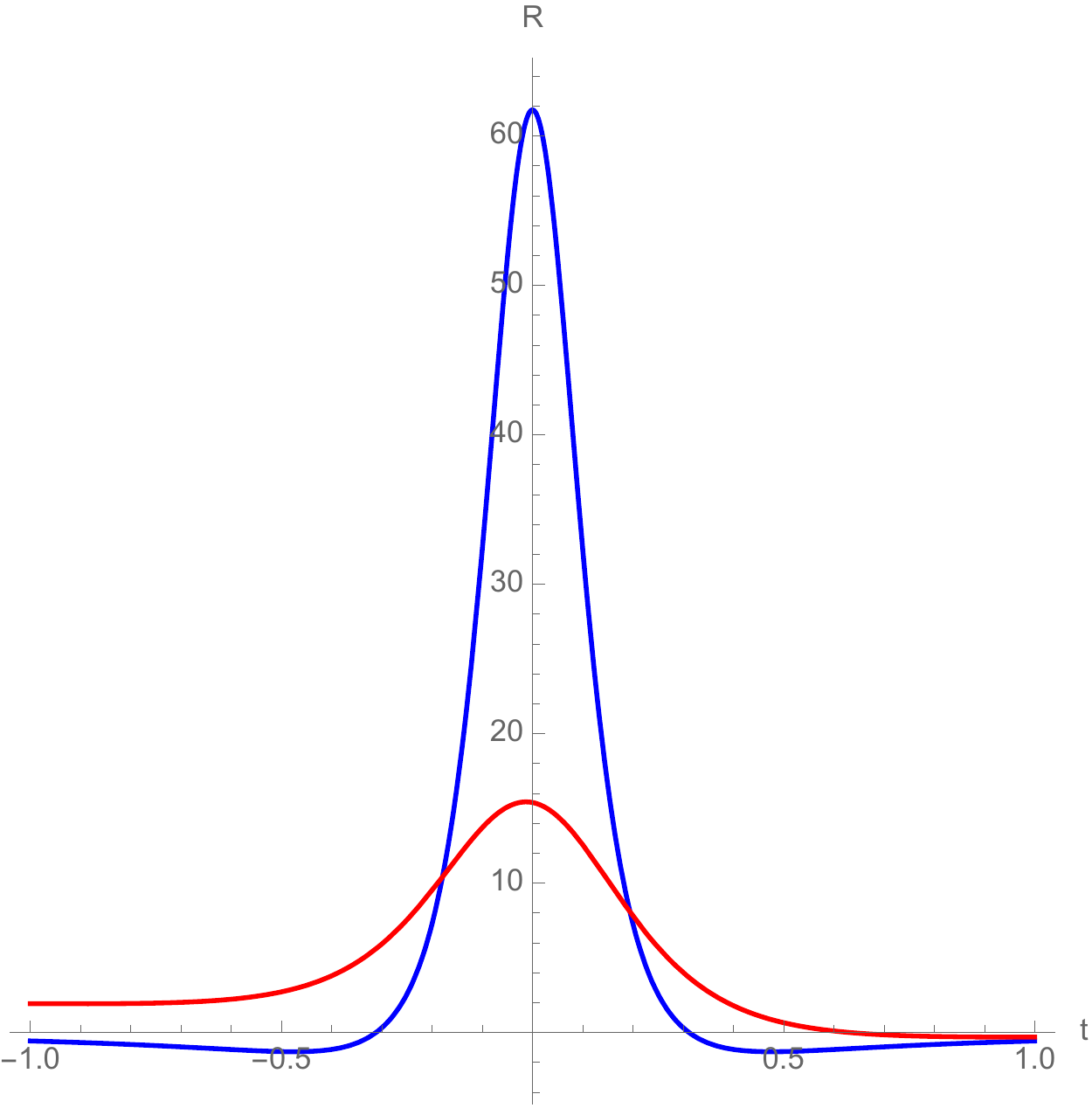}\label{fig2}}
	\hspace{2cm}
	\subfloat[]{\includegraphics[width=7cm, height=5cm]{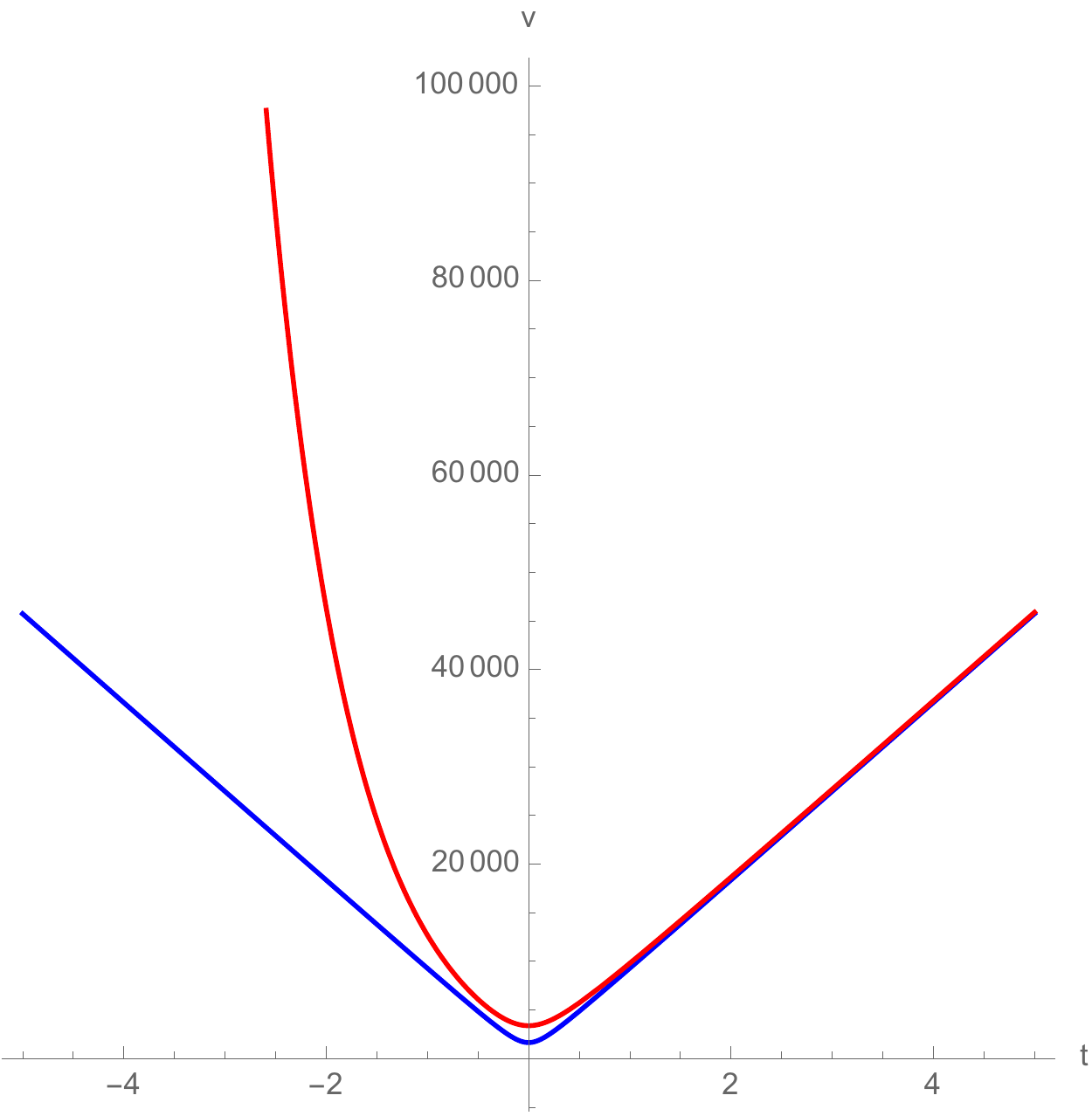}\label{fig3}}
\caption{\small \raggedright The evolution of the curvature $R$ and the volume $v$ during the epoch $\phi \geq 0$ of fig.~\ref{figs1} (as functions of cosmological time $t$) for studied model is compared against that of mainstream LQC. The bounce occurs at $t=0$ as a function of the cosmological time $t$, corresponding to $\phi_0$. Notice that in our model $R$ reaches $0$ in the future classical FLRW phase, but a finite non vanishing value in the past de Sitter phase (a), consistent with the non symmetric bounce shown in (b).}
\end{figure*}

This shows that the $b \rightarrow 0$ limit corresponds to a classically expanding/contracting Universe (we get Friedmann equation with a massless scalar field as matter), while in the $b \rightarrow \pm b_o$ limit we have cosmological constant domination, which produces a de Sitter expanding/contracting phase consistent with the quantum dynamics discussed above. It should not be surprising that quantum gravity effects are present for large volume and low matter energy density, since in the $b \rightarrow \pm b_o$ limit the curvature remains of Planckian order (see fig.\ref{fig2}).

\noindent\textbf{Conclusions:} In this article we studied the effects of an alternative quantization of the Hamiltonian in the context of LQC. This alternative is obtained from a different implementation of the Hamiltonian in the quantum theory based on T.\ Thiemann's regularization \eqref{eq:K}, and which leads to the modified evolution equation \eqref{eq:ev}. 
The Hamiltonian operator in \eqref{eq:ev} admits self-adjoint extensions parametrized by $\beta \in U(1)$; the eigenfunctions of each extension can be constructed numerically and their asymptotic behaviour is known analytically \eqref{eq:e-as}. Having the eigenstates, we can compute the quantum evolution of any physical state. The emerging evolution picture is discussed in detail 
below eq.~\eqref{rhoc}. {To recall, in backward evolution we observe: first a contracting phase following the predictions of GR, ending with a bounce (resolving the classical singularity) and the} transition to a contracting de Sitter phase. This phase is followed by a transition through past scri at $\phi=0$ to an expanding de Sitter phase, which is connected, through another bounce, to a contracting phase approaching the classical solution in the far past.

What is remarkable is that the states remain sharply peaked throughout the entire evolution. Furthermore their trajectories are to high accuracy mimicked by those generated by the effective Hamiltonian in \eqref{Effec-Ham}. In fact, this effective Hamiltonian coincides with the leading order (in a semiclassical expansion) of the expectation value of the Hamiltonian operator on coherent states peaked on flat FRLW in full LQG (see \cite{DL17, *DL17L}). As the effective theory is parametrized by the cosmic time rather than the matter clock, it displays two distinct cases of ``global'' evolution, distinguished by the sign of the value $b_B$ of $b$ at the bounce and corresponding to $\phi<0$ (for $b_B>0$) and $\phi>0$ respectively. The evolution then ends/starts (respectively) with the expanding/contracting de Sitter phase. 
The de Sitter phase is always characterized by an emergent cosmological constant \eqref{eq:Lambda} and large curvature (fig.\ref{fig2}). This result -- obtained via effective dynamics -- is in striking agreement with the quantum evolution discussed above and it modifies the standard bounce paradigm of LQC (see fig.\ref{fig3}).

An interesting aspect of our results is the existence of a transition from expanding to contracting de Sitter epoch, which in fig.~\ref{figs1} happens at $\phi=0$. This issue has already been discussed in \cite{pa-posL}. On one hand, since the de Sitter expanding/contracting Universe with a scalar field is future/past complete, the two sectors $b<0$ and $b>0$ are geodesically complete, thus from the classical spacetime perspective they constitute separate Universes. On the other hand, the trajectories of locally observable quantities (for example matter energy density) as functions of $\phi$ have a unique analytic extension through that point. Therefore, from the quantum theory perspective (where the time problem forced us to use the matter as a clock) the extension of spacetime past the transition point is natural. Such extension can be interpreted as a variant of the cyclic cosmology proposed in \cite{p-cycl}. Unlike there, however, the quantum evolution in our model connects the future scri of one aeon with the past scri of the next one, instead of its singularity. Independently of the considered perspective, the Planckian de Sitter epoch may strongly affect the predictions regarding the structure of perturbations, as many of their modes will remain outside of the horizon through the entire epoch. In turn, this is expected to provide a well defined ``initial'' perturbation power spectrum near the last bounce, determined by the conditions at the de Sitter transition point, which in fig.~\ref{figs1} happens at $\phi=0$, treated as the true point of origin. These questions however remain for future research.
\clearpage

\noindent\textbf{Acknowledgements:}
M.\ A.\ acknowledges the support of the Polish Narodowe Centrum Nauki (NCN) grant 2011/02/A/ST2/00300.  K.\ L.\ thanks the German  National  Merit  Foundation  for their financial support. T.P. acknowledges the support of the Polish Narodowe Centrum Nauki (NCN) grant 2012/05/E/ST2/03308.

\bibliography{adlp-letter}{}
\bibliographystyle{apsrev4-1}

\end{document}